\documentclass[pra,11pt]{revtex4-2}
\usepackage{amsmath}
\usepackage{amssymb}
\usepackage{revsymb}
\usepackage{amsthm}
\usepackage{verbatim}
\usepackage{listings}
\usepackage[pdftex]{graphicx}
\usepackage{scholax}
\usepackage[scaled=1.075,ncf,vvarbb]{newtxmath}
\usepackage[mathscr]{euscript}
\usepackage[T1]{fontenc}
\usepackage[skip=10pt]{parskip}
\usepackage{enumerate}
\usepackage{microtype}
\usepackage{hyperref}
\usepackage[normalem]{ulem}
\usepackage{titlesec}

\newcommand{\mrm}{\mathrm}

\newcommand{\id}{\mathbb{1}}
\newcommand{\ket}[1]{{\left| {#1} \right\rangle}}
\newcommand{\bra}[1]{{\left\langle {#1} \right|}}

\newcommand{\braket}[2]{{\left\langle {#1}|{#2} \right\rangle}}
\newcommand{\avg}[1]{\left\langle{#1}\right\rangle }

\newcommand{\Fvec}{\vec{\mathcal{F}}}

\newcommand{\Fsca}{\mathcal{F}}

\newcommand{\Grad}{\vec{\nabla}}

\titleformat{\section}
{\large\sl}
{\thesection.}{1em}{}

\titleformat{\subsection}
{\sl}
{\thesubsection.}{1em}{}

\titleformat{\subsubsection}
{\normalfont}
{\thesubsubsection.}{1em}{}

\begin{document}
\title{What is a Schiff moment anyway?}
\author{Amar Vutha}
\affiliation{Department of Physics, University of Toronto, Toronto ON M5S 1A7, Canada}
\begin{abstract}
Schiff moments of atomic nuclei are of considerable interest to experiments searching for undiscovered new physics that breaks time-reversal symmetry. I develop a simple picture of the Schiff moment of a charge distribution, and discuss the interaction of the Schiff moment of a nucleus with the electron in an atom.
\end{abstract}

\maketitle
The breakdown of time-reversal (T) symmetry is believed to be an essential missing piece of the puzzle that would explain why there is vastly more matter in the universe than antimatter \cite{Khriplovich2012}. For this reason, a large number of experiments are engaged in looking for evidence of T-symmetry violation \cite{Chupp2019}.
Experimental searches for T-violation that could occur inside atomic nuclei typically focus on a property known as the nuclear Schiff moment \cite{Schiff1963,Engel2025}. An array of ongoing experiments are searching for Schiff moments of various nuclei using precision techniques, in systems as varied as atoms \cite{Graner2016,Bishof2016}, molecules \cite{Grasdijk2021} and solids \cite{Ramachandran2023}. So a nuclear Schiff moment is evidently an important property that has lofty implications for our understanding of fundamental physics. 

But what \emph{is} a Schiff moment? Informally, a Schiff moment is sometimes described as the difference between the mean-squared radii of the charge and the electric-dipole distributions \cite{Wilkening1984,Vandevis2025}. Formal definitions can be found in a number of technical articles, e.g., \cite{Auerbach1996,Spevak1997,Flambaum2002,Liu2007,Flambaum2012} -- however, these are often written in the sophisticated language of theoretical nuclear physics, which can obscure the simple electrostatics underlying the notion of a Schiff moment. 

The purpose of this note is to provide a simple self-contained introduction to the concept of a Schiff moment of a charge distribution, starting from undergraduate-level electrostatics and quantum mechanics. I hope that it may be useful to undergraduates or graduate students who are interested in precision searches for physics beyond the Standard Model of particle physics. Students and instructors of electromagnetism courses might also be interested in this topic, as it furnishes an example from basic electrostatics that is connected to contemporary research. Readers interested in an alternative introduction to the Schiff moment are referred to Section 1.2 of Ref.\ \cite{Kastelic2024}.

\section{Electron interaction with a nuclear charge distribution}
The potential energy for the electrostatic interaction between an electron (position $\vec{r}$, charge  $-e$) and a nuclear charge distribution (position $\vec{R}$, charge density $\rho_n$) is
\begin{equation}
V_{en}(\vec{r}) = \frac{-e}{4 \pi \epsilon_0} \int d^3R \, \frac{\rho_n(\vec{R})}{|\vec{r} - \vec{R}|}.
\end{equation}
The integral $\int d^3R$ is over the volume of the nucleus. To break up this interaction into simpler parts, consider the Taylor expansion of $1/|\vec{r} - \vec{R}|$ that appears inside the integral. 
\begin{equation}
\frac{1}{|\vec{r} - \vec{R}|} = \frac{1}{r}  -X_i \, \nabla_i \frac{1}{r} + \frac{1}{2!} X_i X_j \, \nabla_j \nabla_i \frac{1}{r} - \frac{1}{3!} X_i X_j X_\ell  \, \nabla_\ell \nabla_j \nabla_i \frac{1}{r}  + \ldots
\end{equation}
Here $X_i$ are the components of $\vec{R}$, $x_i$ are the components of $\vec{r}$ and $\nabla_i = \frac{\partial }{\partial x_i}$. The cartesian indices $i,j,\ell$ range from 1 to 3. Repeated indices are summed over. Mathematically, the Taylor series above is convergent when $R < r$. Physically, this means the above expansion is valid when the electron is located outside the extent of the nuclear charge distribution, a reasonable approximation since the length scale of an electron wavefunction in an atom ($\sim 10^{-10}$ m) is much larger than the typical size of a nucleus ($\sim 10^{14}$ m) \footnote{Nevertheless, atomic electrons do have a small nonzero probability of being found \emph{inside} the nucleus of an atom. This effect leads to a small correction to the simple picture developed discussed in this note. Readers interested in the details of such corrections can find them in specialized papers such as Ref.\ \cite{Flambaum2001}.}.

The Taylor expansion allows us to separate the electron coordinate and its derivatives from the integral over $d^3R$. As a result, the electron-nucleus interaction conveniently splits into a sum of terms, each of which is a product of a nuclear quantity and an electronic quantity. Up to 3rd order in $\vec{R}$, the interaction potential is $V_{en} = V_{E0} + V_{E1} + V_{E2} + V_{E3}$, where
\begin{equation}
\begin{split}\label{eq:multipole_expansion}
V_{E0} & =  k \left(\int d^3R \, \rho_n \right) \left[ \frac{1}{r} \right]  \\
V_{E1} & =  - k \left(\int d^3R \, \rho_n \, X_i \right)  \left[ \frac{- x_i}{r^3} \right] \\
V_{E2} & = \frac{k}{2} \left(\int d^3R \, \rho_n \, X_i X_j \right) \left[ -\frac{4\pi}{3} \delta_{ij}\, \delta(\vec{r}) + \frac{3 x_i x_j - r^2 \delta_{ij}}{r^5}    \right] \\
V_{E3} & = - \frac{k}{6} \left(\int d^3R \, \rho_n \, X_i X_j X_\ell \right) \left[  -\frac{4\pi}{5} \delta_{\{ij} \, \nabla_{\ell\}} \delta(\vec{r}) +   \frac{ 3 \delta_{\{ij} x_{\ell\}} r^2 - 15 x_i x_j x_\ell }{r^7}  \right]. 
\end{split}
\end{equation}
Here $k = -e/4 \pi \epsilon_0$ for ease of notation. The $E0,E1,E2$ and $E3$ interactions are traditionally called the electric monopole, dipole, quadrupole and octupole interactions respectively. Each of these is a product of a nuclear multipole moment (in round brackets) that depends on $\vec{R}$, and a field generated by the electron (in square brackets) that depends on $\vec{r}$. The gradients of $1/r$ appearing above are derived in the Appendix using a simple regularization method. The symmetrized 3rd rank tensor formed from the 2nd rank tensor $A_{ij}$ and the vector $B_\ell$ is compactly written as $A_{\{ij} B_{\ell\}} = A_{ij} B_\ell + A_{\ell i} B_j + A_{j\ell} B_i$. 

Within the $E2$ and $E3$ interaction potentials, there are two types of electronic terms. The so-called ``contact'' terms are the ones containing $\delta(\vec{r})$ or its derivatives \cite{Gray2009,Gray2010}, whereas the non-contact terms involve traceless tensors formed from components of $\vec{r}$. 
Let us focus our attention on the contact $E3$ term 
\begin{equation}
\begin{split}\label{eq:cE3}
V_{cE3} & = -\frac{e}{30 \epsilon_0} \left( \int d^3R \,  \rho_n(\vec{R}) \, X_i X_j X_\ell \right) \left[ (\delta_{ij} \nabla_\ell + \delta_{j\ell} \nabla_i + \delta_{\ell i} \nabla_j ) \delta(\vec{r}) \right]   \\  
& =  -\left( \frac{1}{10}\int d^3R \, \rho_n(\vec{R}) \, R^2 \, \vec{R} \right) \cdot \left[ \frac{e}{\epsilon_0} \Grad \delta(\vec{r}) \right]  \\
& = -\vec{\mathscr{S}}_0 \cdot \Fvec.
\end{split}
\end{equation}
The quantity $\vec{\mathscr{S}}_0 = \frac{1}{10}\int d^3R \, \rho_n(\vec{R}) \, R^2 \, \vec{R}$ is the first piece of the Schiff moment of the charge distribution $\rho_n(\vec{R})$. I have also defined $\Fvec = \frac{e}{\epsilon_0} \Grad \delta(\vec{r})$, the vector that couples to the nuclear Schiff moment -- this will be discussed in more detail in Section \ref{sec:F_vector}.

But there is more to the story: we need to consider one more effect for a nucleus in an atom.

\section{Shielding of the nuclear electric dipole moment in an atom}
{An electric dipole moment of a charge distribution that also has nonzero total charge is equivalent to a displacement of the centre of charge.} Inside an atom, this fact has an interesting consequence.

Let the nucleus have a nonzero electric dipole moment (EDM) $\vec{D} = \int d^3R \, \rho_n \vec{R}$. The combination of its electric charge $Q = \int d^3R	 \rho_n$ (monopole moment) and dipole moment implies that the centre of charge of the nucleus is at $\Delta \vec{R} = \frac{\vec{D}}{Q}$.

In atomic physics, electron wavefunctions are usually calculated with respect to an origin located at the {centre of charge} of the nucleus. Therefore, physically displacing the nuclear centre of charge by a small amount $\Delta \vec{R}$ -- sufficiently slowly for the adiabatic theorem \cite{Kato1950} to apply --
just displaces all the electron eigenfunctions by $\Delta \vec{R}$ as well. This operation is mathematically the same as shifting the origin to the centre of charge of the nucleus, which is equivalent to replacing $\vec{r}$ with $\vec{r} + \Delta \vec{R}$ in Equation (\ref{eq:multipole_expansion}).

The modified electron-nuclear interaction (which I will denote as $V_{en}'$), up to first order in the displacement $\Delta \vec{R}$, is
\begin{equation}\label{eq:first_order_displacement}
V'_{en}(\vec{r}) = V_{en}(\vec{r} + \Delta \vec{R}) = V_{en}(\vec{r}) + \frac{D_i}{Q}\nabla_i V_{en}(\vec{r}) = V_{E0}' + V_{E1}' + V_{E2}' + V_{E3}',
\end{equation}
where the modified multipole interactions are 
\begin{equation}
\begin{split}
V'_{E0} & = k \left(\int d^3R \, \rho_n \right) \left[ \frac{1}{r} \right]  \\
V'_{E1} & = -k \left(\int d^3R \, \rho_n X_i \right)  \left[ -\frac{x_i}{r^3} \right]  + \frac{D_j}{Q}  \nabla_j V_{E0} \\
V'_{E2} & = \frac{k}{2} \left(\int d^3R \, \rho_n \, X_i X_j \right) \left[ -\frac{4\pi}{3} \delta_{ij}\, \delta(\vec{r}) + \frac{3 x_i x_j - r^2 \delta_{ij}}{r^5}    \right] +  \frac{D_i}{Q}  \nabla_i V_{E1} \\
V'_{E3} & = - \frac{k}{6} \left(\int d^3R \, \rho_n \, X_i X_j X_\ell \right) \left[  -\frac{4\pi}{5} \delta_{\{ij} \nabla_{\ell\}} \delta(\vec{r}) +   \frac{ 3 \delta_{\{ij} x_{\ell\}} r^2 - 15 x_i x_j x_\ell }{r^7}  \right] + \frac{D_i}{Q}  \nabla_i V_{E2}.
\end{split}
\end{equation}

First, consider the $E1$ interaction, which becomes
\begin{equation}
\begin{split}
V'_{E1} & = -k \left(\int d^3R \, \rho_n X_i \right)  \left[ -\frac{x_i}{r^3} \right]  + k \frac{D_j}{Q} \left(\int d^3R \, \rho_n \right)   \left[ -\frac{x_j}{r^3} \right] = 0,
\end{split}
\end{equation}
since $Q = \int d^3R \, \rho_n$ and $D_i = \int d^3R \, \rho_n X_i$. The EDM of the nucleus is shielded by the rearrangement of the electron charge distribution, leading to zero measurable contribution of the EDM to the $E1$ interaction. This is Schiff's shielding theorem \cite{Schiff1963}.

Next consider the contact part of the modified $E3$ interaction, which acquires a correction from the original contact $E2$ interaction to become
\begin{equation}\label{eq:schiff_moment_interaction}
\begin{split}
V'_{cE3} & =  -\left( \frac{1}{10} \int d^3R \, \rho_n R^2 X_i \right) \cdot \left[ \frac{e}{\epsilon_0} \nabla_i \delta(\vec{r}) \right]  
+  \frac{e}{6 \epsilon_0} \frac{D_j}{Q} \left(\int d^3R \, \rho_n R^2\right)   \left[ \nabla_j \delta(\vec{r}) \right] \\
& = - \left( \frac{1}{10} \int d^3R \, \rho_n R^2 X_i  - \frac{1}{6} \frac{D_i}{Q} \int d^3R \, \rho_n R^2 \right)  \left[ \frac{e}{\epsilon_0} \nabla_i \delta(\vec{r}) \right]  \\
& = -(\vec{\mathscr{S}}_0 + \vec{\mathscr{S}}_1) \cdot \Fvec.
\end{split}
\end{equation}
The combination 
\begin{equation}\label{eq:schiff_moment_def}
\vec{\mathscr{S}} = \vec{\mathscr{S}}_0 + \vec{\mathscr{S}}_1 = \frac{1}{10} \int d^3R \, \rho_n(\vec{R}) R^2 \vec{R}  - \frac{1}{6} \frac{\vec{D}}{Q} \int d^3R \, \rho_n(\vec{R}) R^2
\end{equation}
is the Schiff moment of the nucleus, as it is usually defined. It consists of the nuclear moment $\vec{\mathscr{S}}_0$, defined in Equation (\ref{eq:cE3}), plus a correction, $\vec{\mathscr{S}}_1 \propto \vec{D}$, due to the displacement of the centre of charge.

There is a further contact term in $V'_{cE3}$ generated by the gradient of the non-contact $\frac{3 x_i x_j - r^2 \delta_{ij}}{r^5}$ term in $V_{E2}$. When this gradient 
is evaluated using the regularization method in the Appendix and simplified with the help of Equation (\ref{eq:grad_delta}), the additional contribution to $V'_{cE3}$ is 
\begin{equation}\label{eq:S2}
-\vec{\mathscr{S}}_2 \cdot \Fvec = \frac{1}{15} \frac{D_j}{Q} \mathscr{Q}_{ij} \left[ \frac{e}{\epsilon_0} \nabla_i \delta(\vec{r}) \right], 
\end{equation}
where $\mathscr{Q}_{ij} = \int d^3R \, \rho_n (3 X_i X_j - X^2 \delta_{ij})$ is the traceless electric quadrupole moment tensor of the nuclear charge distribution, and the components of the vector $\vec{\mathscr{S}}_2$ are
\begin{equation}\label{eq:S2}
{\mathscr{S}}_{2,i} = \frac{-1}{15} \frac{D_j}{Q}\mathscr{Q}_{ij}.
\end{equation}
Some papers in the literature (e.g., \cite{Flambaum2012}) include this term and define the Schiff moment as $\vec{\mathscr{S}} = \vec{\mathscr{S}}_0 + \vec{\mathscr{S}}_1 + \vec{\mathscr{S}}_2$. The $\vec{\mathscr{S}}_2$ term is relevant for nuclei that have electric quadrupole moments.

The Schiff moment $\vec{\mathscr{S}}$ is purely a property of the distribution $\rho_n(\vec{R})$. Some of its properties are:
\begin{enumerate}[(i)]
\item $\vec{\mathscr{S}}$ can be defined for any charge distribution, even in classical electromagnetism. There is nothing intrinsically nuclear or quantum-mechanical about it.

\item $\vec{\mathscr{S}}$ has units of charge $\times$ length$^3$ (C m$^3$ in SI, or $e$ fm$^3$ in units appropriate for nuclei).

\item The Schiff moment is a vector quantity. But it is \emph{not} the EDM: the contact $E3$ interaction is not the same as the $E1$ interaction. 
\end{enumerate}

In quantum mechanics, the Schiff moment $\vec{\mathscr{S}}$ is a vector operator in the space of nuclear states. In a nucleus with spin $\vec{I}$, the expectation value of the Schiff moment vector in any nuclear state $\ket{\Psi}$ is parallel to the expectation value of the angular momentum vector, $\avg{\vec{\mathscr{S}}} = \bra{\Psi} \vec{\mathscr{S}} \ket{\Psi} \propto \bra{\Psi} \vec{I} \ket{\Psi}$. This property follows from the Wigner-Eckart theorem \cite{Sakurai}. Note, however, that $\vec{I}$ is an operator that is odd under time-reversal (T) symmetry: a spinning nucleus with angular momentum $-\vec{I}$ can be mapped to a nucleus with angular momentum $\vec{I}$ whose dynamics is time-reversed. Therefore, if $\avg{\vec{\mathscr{S}}} \neq 0$ in an angular momentum eigenstate of a nucleus, then the nuclear charge distribution is odd under T-symmetry. Which means that the nucleus spinning in one sense has a \textit{different} charge distribution than the nucleus spinning in the opposite sense! Such an unusual charge distribution can only arise (as far as we know) from new particles or interactions that break T-symmetry inside nuclei. This is why a nuclear Schiff moment is a useful target of experimental searches, since it is an indicator of T-violating physics originating from beyond the Standard Model.
	
\section{Charge distribution with a Schiff moment}
\begin{figure}[h!]
\centering
\includegraphics[width=0.9\columnwidth]{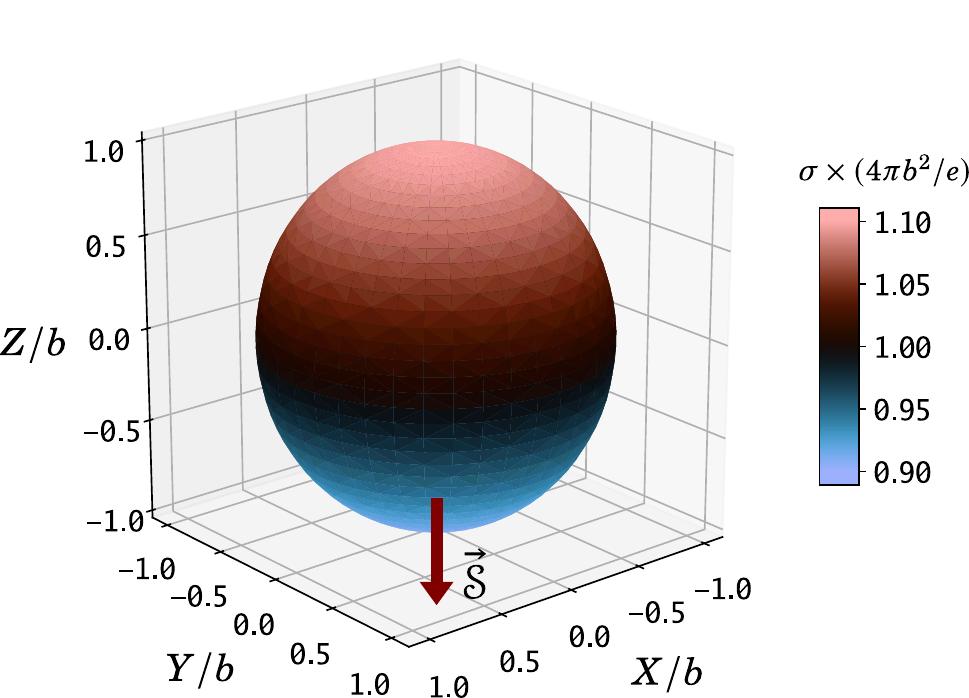}
\caption{An example of a charge distribution with a nonzero Schiff moment. The color bar shows the surface charge density $\sigma(\theta,\phi) \propto (1 + \eta \cos \theta)$ on the sphere, with $\eta = 0.1$. The direction of the Schiff moment of this charge distribution is indicated by the arrow. }\label{fig:schiff_moment_sphere}
\end{figure}

Let us construct a concrete example of a charge distribution with a nonzero Schiff moment, to observe how the definitions above can be applied. Consider a charge distribution confined to the surface of a sphere of radius $b$. Let the charge per unit area on the surface be $\sigma(\theta,\phi) = \frac{e}{4\pi b^2} \left(1 + \eta \cos \theta \right)$, where $\theta,\phi$ are the usual spherical polar angles. The surface charge density is illustrated in Figure \ref{fig:schiff_moment_sphere}. The charge distribution is azimuthally symmetric, and therefore only the $z$-components of its vector moments are nonzero. The relevant nonzero moments of this distribution are
\begin{equation}
\begin{split}
\int d^3R \, \rho(\vec{R}) & \to b^2 \int_0^{2\pi} d\phi \int_0^\pi \sin\theta \, d\theta \, \sigma(\theta,\phi) =  e \\
\int d^3R \, \rho(\vec{R}) \, Z & \to b^3 \int_0^{2\pi} d\phi \int_0^\pi \sin\theta \, d\theta \, \sigma(\theta,\phi) \, \cos \theta =  \frac{1}{3} \eta eb \\
\int d^3R \, \rho(\vec{R}) \, R^2 & \to b^4 \int_0^{2\pi} d\phi \int_0^\pi \sin\theta \, d\theta \, \sigma(\theta,\phi) =  eb^2 \\
\int d^3R \, \rho(\vec{R}) \, R^2 Z& \to b^5 \int_0^{2\pi} d\phi \int_0^\pi \sin\theta \, d\theta \, \sigma(\theta,\phi) \, \cos \theta =  \frac{1}{3}\eta eb^3. 
\end{split}
\end{equation}

Since this distribution has both a monopole and a dipole moment, the centre of charge is not located at the centre of the sphere, but at $\Delta \vec{R} = \frac{\eta b}{3} \, \hat{z}$ instead. 

Let us assume the dimensionless parameter $\eta \ll 1$, so that the displacement of the centre of charge $\Delta \vec{R}$ is small compared to $b$, and we are in the regime described by Equation (\ref{eq:first_order_displacement}). Then the $z$-component of $\vec{\mathscr{S}}$ is easily calculated using Eq.\ (\ref{eq:schiff_moment_def}) to be $\mathscr{S}_Z = \frac{1}{30}\eta e b^3 - \frac{1}{18}\eta e b^3 = -\frac{1}{45}\eta e b^3$. 

It is interesting to observe that the Schiff moment of this charge distribution is in the \emph{opposite} direction to its electric dipole moment, which may not be what you intuitively expected at first.

\section{The $\Fvec$ vector}\label{sec:F_vector}
Having seen how the definition of the Schiff moment arises naturally from the electron-nucleus electrostatic interaction, let us explore the physics of the electronic quantity $\Fvec$ that couples to the nuclear Schiff moment. The definition of $\Fvec$ in Equation (\ref{eq:schiff_moment_interaction}), which involves the gradient of a delta-function, is a rather singular mathematical object. To gain physical understanding without getting lost in mathematical subtleties, let us replace the $\delta(\vec{r})$ in the definition of $\Fvec$ with its smooth regularized version described in the Appendix, using a regularization length scale $a$. The gradient of the delta-function is then
\begin{equation}
\begin{split}
\Grad \delta(\vec{r}) & \approx \frac{3}{4 \pi a^3} \Grad \left(1 + \frac{r^2}{a^2} \right)^{-5/2} \\
& = -\frac{15}{4 \pi a^5}\frac{ \vec{r}}{\left(1 + \frac{r^2}{a^2} \right)^{7/2}} \\
& = - \frac{\vec{r} \, g(r)}{a^5},
\end{split}
\end{equation}
where the dimensionless function $g(r) = \frac{15}{4\pi} \left(1 + \frac{r^2}{a^2} \right)^{-7/2}$ is defined for convenience. Thus a non-singular approximation of $\Fvec$ generated by the electron is $\Fvec = -\frac{e \vec{r} \, g(r)}{\epsilon_0 a^5}$. Fig.\ 1 shows an illustration of the $\Fvec$ vector field produced by an electron in a classical orbit around a nucleus.

\begin{figure}\label{fig:F_vector}
\centering
\includegraphics[width=0.5\columnwidth]{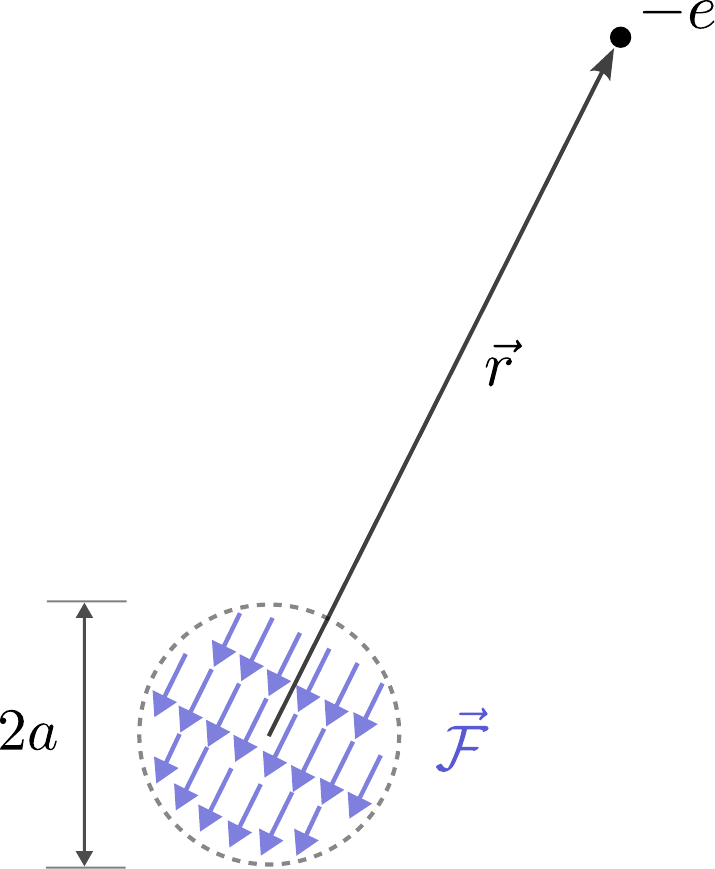}
\caption{Illustration of the vector field $\Fvec$ produced by an electron in an atom. This vector field is localized and approximately constant within a sphere of radius $\sim a$ (which we can take to be much smaller than the size of an electron orbit), and is directed opposite to the electron's position vector $\vec{r}$. The vector $\Fvec$ couples to the Schiff moment of the nucleus, as described by Equation (\ref{eq:schiff_moment_interaction}).}
\end{figure}

But electrons do not move classically inside atoms. It is instructive to calculate the quantum mechanical expectation value of the $\Fvec$ operator for an electron in a simple hydrogenic atom. Without loss of generality, we can choose a coordinate system that is oriented along $\Fvec$, so let us simply calculate the expectation value of the $z$ component of $\Fvec$. First consider an electron in an $s$ state (orbital angular momentum $\ell=0$) with principal quantum number $n$. It turns out that $\bra{ns}\Fsca_z\ket{ns} = 0$. This is a consequence of two facts: i) the operator $\Fsca_z$ is a component of a polar vector, and ii) angular momentum eigenstates like have definite parity. Denoting the unitary parity-reversal operator by $P$, these facts are expressed by the relations $P \Fsca_z P^\dagger = -\Fsca_z$ and $P \ket{ns} = + \ket{ns}$. And so, making use of the unitary property $P^\dagger P = \id$, 
\begin{equation}\label{eq:parity_zero}
\bra{ns}\Fsca_z\ket{ns} = \bra{ns}P^\dagger P \, \Fsca_z \, P^\dagger P\ket{ns} = ( \bra{ns} P^\dagger) \, (-\Fvec_z) \left( P \ket{ns} \right) = -\bra{ns}\Fsca_z\ket{ns} = 0.
\end{equation}
Consider an electron in the state $\ket{n' p_z}$ instead, where $n'$ is the principal quantum number, and $p_z$ denotes the orbital with angular momentum $\ell=1$ and azimuthal quantum number $m_\ell=0$. Here too, since $P \ket{p_z} = -\ket{p_z}$, it is easy to verify that $\bra{n' p_z}\Fsca_z\ket{n' p_z} = 0$, following the same lines as Equation (\ref{eq:parity_zero}). So far, it seems like neither $s$ states nor $p$ states generate any interesting $\Fvec$ for a nuclear Schiff moment to interact with.

Now consider an electron in a superposition of $ns$ and $n'p_z$ states, $\ket{\psi} = c_s \ket{ns} + c_p \ket{n'p_z}$, with complex coefficients $c_s$ and $c_p$, and let us evaluate $\bra{\psi} \Fsca_z \ket{\psi}$. Since $\bra{ns}\Fsca_z\ket{ns}$ and $\bra{n' p_z}\Fsca_z\ket{n' p_z}$ are both equal to zero, only the cross terms between $\bra{ns}$ and $\ket{n'p_z}$ contribute to the expectation value: $\bra{\psi} \Fsca_z \ket{\psi} = 2 \, \mrm{Re}(c_p^* c_s) \bra{n'p_z} \Fsca_z \ket{ns}$. Let us define $\xi =  \mrm{Re}(c_p^* c_s)$ for compactness of notation. To evaluate the matrix element $\bra{n'p_z} \Fsca_z \ket{ns}$, we need the spatial wavefunctions of the atomic states, which are $\braket{\vec{r}}{ns} = \mathfrak{R}_{ns}(r) Y_{00}(\theta \phi)$ and $\braket{\vec{r}}{n'p_z} = \mathfrak{R}_{n'p}(r) Y_{10}(\theta \phi)$, where $\mathfrak{R}_{n\ell}(r)$ are hydrogenic radial functions and $Y_{\ell m}(\theta \phi)$ are spherical harmonics. Therefore 
\begin{equation}
\begin{split}
\bra{\psi} \Fsca_z \ket{\psi} & = 2 \xi \bra{n'p_z} \Fsca_z \ket{ns} \\
& = -\frac{2 \xi e }{\epsilon_0 a^5} \int_0^\infty 4 \pi r^2 dr\, \mathfrak{R}_{ns}(r)  \mathfrak{R}_{n'p}(r) \, r g(r) \int_0^{2\pi} d\phi \int_0^\pi d\theta \, \sin\theta \, Y_{00}(\theta \phi) \cos \theta \, Y_{10} (\theta \phi) \\
& =  -\frac{2 \xi e }{\sqrt{3} \epsilon_0 a^5}  \int_0^{r_1} 4 \pi r^2 dr \, \mathfrak{R}_{ns}(r)  \mathfrak{R}_{n'p}(r) \, r g(r).
\end{split}
\end{equation}
In the last line, I also used the fact that $g(r)$ drops rapidly to zero when $r$ is much larger than $a$. So it is sufficient to take the radial integral to an upper limit $r_1 \gtrsim a$ (say $r_1 = 5a$), instead of $\infty$. The exact value of $r_1$ will turn out to be unimportant.
 
Let us set the regularization length scale $a$ (which, recall is just a mathematical artifice) to be much smaller than the typical size of an electron orbital in an atom. For a hydrogenic atom with nuclear charge $Z$, the size of atomic orbitals is $\sim a_0/Z$, where $a_0$ is the Bohr radius. Setting $a \ll a_0/Z$, inside a sphere of radius $a$ we can approximate the wavefunctions by their leading-order Taylor expansion near the origin, $\mathfrak{R}_{ns}(r) \approx \alpha$ and $\mathfrak{R}_{n'p}(r) = \beta r$. The constants $\alpha$ and $\beta$ are easily obtained from tables of hydrogenic wavefunctions, and their values are $\alpha = 2 \sqrt{\frac{Z^3}{n^3 a_0^3}}$ and $\beta = \frac{2Z}{3 a_0} \sqrt{\frac{Z^3}{n'^3 a_0^3} \left(1 - \frac{1}{n'^2}\right)}$.

Thus the $z$-component of the $\Fvec$ vector field produced by an electron in state $\ket{\psi}$ is
 \begin{equation}
 \begin{split}
 \bra{\psi} \Fsca_z \ket{\psi} =  -\frac{2 \xi e }{\sqrt{3} \epsilon_0 a^5} \alpha \beta \int_0^{r_1} 4 \pi r^2 dr \, r^2 g(r). 
 \end{split}
\end{equation}
Remarkably, the integral $\int_0^{r_1} 4\pi r^2 dr\,r^2 g(r) = 3 a^5$ whenever $r_1 \gtrsim a$, regardless of the exact value of $r_1$, since the step-like function $g(r)$ only has appreciable support over a region of radius $\sim a$ around the origin. Hence $\bra{\psi}\Fsca_z\ket{\psi}$ has a finite and well-defined value, regardless of our choice of specific values for the regularization length scale $a$ and the integral limit $r_1$, so long as $a \ll r_1 \ll a_0/Z$. All of which leads to the expression
\begin{equation}
\bra{\psi} \Fsca_z \ket{\psi} =  -\frac{2 \sqrt{3} \xi e }{\epsilon_0} \alpha \beta  = -\frac{8 \xi}{\sqrt{3}} \frac{ e Z^4}{\epsilon_0 a_0^4} \sqrt{\frac{n'^2 - 1}{n^3 n'^5}}.
\end{equation}
Note that the expectation value of $\Fsca_z$ goes to zero if either $c_s$ or $c_p$ is zero, so only an electron in a superposition of $s$ and $p$ states generates nonzero $\Fvec$ in the vicinity of the nucleus. Therefore, electrically-polarized atoms or polar molecules, wherein electron eigenfunctions consist of linear combinations of $s$ and $p$ orbitals, are necessary in order to generate an $\Fvec$ vector field to couple to a nuclear Schiff moment. Also note that $\bra{\psi} \Fsca_z \ket{\psi} \propto Z^4$ is a steeply rising function of $Z$. These are the reasons why current experiments searching for nuclear Schiff moments use high-$Z$ nuclei contained within electrically-polarized atoms or molecules (e.g., \cite{Graner2016,Bishof2016,Grasdijk2021}).

When a nucleus in state $\ket{\Psi}$ interacts with an electron in state $\ket{\psi}$, then the energy shift of the combined atomic state $\ket{\Psi}\ket{\psi}$ due to the Schiff moment interaction is
\begin{equation}
\Delta E = \bra{\psi} \bra{\Psi} \, V_{cE3}' \, \ket{\Psi} \ket{\psi}  = - \bra{\Psi}\mathscr{S}_z\ket{\Psi} \, \bra{\psi} \Fsca_z \ket{\psi}
\end{equation}
to first order in perturbation theory. If the nucleus is prepared in the spin eigenstate $\ket{\Psi} = \ket{I,m_I}$, the expectation value of the nuclear Schiff moment operator is 
\begin{equation}
\bra{\Psi}\mathscr{S}_z\ket{\Psi} = \bra{I,m_I} \mathscr{S}_z \ket{I,m_I} = \frac{ |\vec{\mathscr{S}}| }{ |\vec{I}| } \bra{I,m_I} I_z \ket{I,m_I} = \frac{ |\vec{\mathscr{S}}| }{ |\vec{I}| } m_I,
\end{equation}
due to the Wigner-Eckart theorem \cite{Sakurai}. The resulting characteristic energy shift of the atom, $\Delta E \propto m_I \, \bra{\psi} \Fsca_z \ket{\psi}$, is measurable in experiments (see, e.g., \cite{Graner2016,Bishof2016}).

In summary, a Schiff moment is a property of a charge distribution that combines aspects of its electric monopole, dipole, quadrupole and octupole moments. The definition of a nuclear Schiff moment, and the physics of its interaction with electrons in atoms, follow from elementary electrostatics and quantum mechanics. 

\subsection*{Acknowledgments}
I benefited from discussing this topic with Mohit Verma, William Zheng, Wesley Campbell, Andrew Jayich and David DeMille. Daniel Comparat and Subham Mahapatra pointed out the $\mathscr{S}_2$ term defined in Equation (\ref{eq:S2}). This work was supported by NSERC and the Canada Research Chairs program.

\newpage
\appendix
\section*{Appendix: Gradients of $1/r$}

Higher-order gradients of the $1/r$ function are rather singular mathematical quantities \cite{Gray2009,Gray2010}. To make their behaviour smoother and more physical, let us consider a ``regularized'' version of $1/r$, which is $\frac{1}{\sqrt{r^2 + a^2}}$. This function is finite and its gradients can be sensibly evaluated, before eventually taking the limit  $a \to 0$.

The first-order gradient of $1/r$ can be obtained in a straightforward manner and contains no surprises. 
\begin{equation}
\lim_{a \to 0} \nabla_i \frac{1}{\sqrt{r^2 + a^2}} =  -\frac{x_i}{(r^2 + a^2)^{3/2}} \longrightarrow -\frac{x_i}{r^3}.
\end{equation}

The second-order gradient is
\begin{equation}
\begin{split}
\nabla_j \nabla_i \frac{1}{\sqrt{r^2 + a^2}} & = -\frac{\delta_{ij}}{(r^2 + a^2)^{3/2}}+ \frac{3 x_i x_j}{(r^2 + a^2)^{5/2}} \\
& = - \frac{\delta_{ij}\, a^2}{(r^2 + a^2)^{5/2}} + \frac{3 x_i x_j - r^2 \delta_{ij}}{(r^2 + a^2)^{5/2}}.
\end{split}
\end{equation}
As $a \to 0$, the second term tends to $(3 x_i x_j - r^2 \delta_{ij})/r^5$ . But the first term needs some care, especially when $r \ll a$. In this regime, the first term approaches $- \delta_{ij} \frac{1}{a^3}$. Its volume integral, over any sphere centred at the origin whose radius is $\gg a$, is equal to $-\frac{4 \pi}{3}{\delta_{ij}}$ regardless of the value of $a$. This fact can be observed from the graph of $\frac{a^2}{(r^2 + a^2)^{5/2}}$, which is a flat-topped function of width $a$ that falls off rapidly, so that the volume of the sphere over which this function is non-negligible approaches $\frac{4\pi}{3}a^3$. Thus its integral over \textit{any volume} sufficiently larger than $a^3$ centred on the origin is equal to $\frac{4\pi}{3}$. In other words,
\begin{equation}
\int_0^\infty \frac{a^2}{(r^2 + a^2)^{5/2}} \, 4\pi r^2 dr = 4\pi \int_0^\infty \frac{t^2 \, dt}{(1+ t^2)^{5/2}} = 4 \pi \left[\frac{t^3}{3(1 + t^2)^{3/2}}\right]^\infty_0 = \frac{4 \pi}{3}. 
\end{equation}
Since the function $\frac{3}{4\pi} \frac{a^2}{(r^2 + a^2)^{5/2}}$ only has support over a sphere of radius $\sim a$, and has a volume integral equal to 1 for all values of $a$, 
\begin{equation}\label{eq:regularized_delta}
\lim_{a \to 0}  \frac{3}{4 \pi a^3} \left( 1 + \frac{r^2}{a^2} \right)^{-5/2} \to \delta(\vec{r}).
\end{equation}
Therefore the second-order gradient of $\frac{1}{r}$ is
\begin{equation}
\nabla_j \nabla_i \frac{1}{r} = -\frac{4\pi}{3} \delta_{ij} \, \delta(\vec{r})  + \frac{3 x_i x_j - r^2 \delta_{ij}}{r^5}.
\end{equation}
The first tensor on the right-hand side has nonzero trace, whereas the second tensor is traceless.

The third-order gradient is
\begin{equation}
\begin{split}\label{eq:triple_gradient_regularized}
\nabla_\ell \nabla_j \nabla_i \frac{1}{\sqrt{r^2 + a^2}} & = \frac{3 \left( \delta_{ij} x_\ell + \delta_{j\ell} x_i + \delta_{\ell i} x_j \right)}{(r^2 + a^2)^{5/2}} - \frac{15 x_i x_j x_\ell}{(r^2 + a^2)^{7/2}} \\
& = \frac{3 \left( \delta_{ij} x_\ell + \delta_{j\ell} x_i + \delta_{\ell i} x_j \right) a^2}{(r^2 + a^2)^{7/2}} + \frac{3 \left( \delta_{ij} x_\ell + \delta_{j\ell} x_i + \delta_{\ell i} x_j \right) r^2 - 15 x_i x_j x_\ell}{(r^2 + a^2)^{7/2}}.
\end{split}
\end{equation}
As $a \to 0$, the second term tends to $\left( 3 \delta_{\{ij} x_{\ell\}} r^2 - 15 x_i x_j x_\ell \right)/ r^7$ . The first term again needs some care. From Equation (\ref{eq:regularized_delta}), 
\begin{equation}\label{eq:grad_delta}
\lim_{a \to 0} -\frac{15}{4 \pi a^5} \, x_\ell \,  \left(1 + \frac{r^2}{a^2} \right)^{-7/2} \to \nabla_\ell \delta(\vec{r}),
\end{equation}
so that the first term in Equation (\ref{eq:triple_gradient_regularized}) tends to $-\frac{4\pi}{5} \delta_{\{ij} \nabla_{\ell\}} \delta(\vec{r})$. Therefore
\begin{equation}
\nabla_\ell \nabla_j \nabla_i \frac{1}{r} = -\frac{4\pi}{5} \delta_{\{ij} \nabla_{\ell\}} \delta(\vec{r}) +   \frac{ 3 \delta_{\{ij} x_{\ell\}} r^2 - 15 x_i x_j x_\ell }{r^7}.
\end{equation}
Once again, the first tensor on the right-hand side has nonzero trace, whereas the second tensor is traceless.

\bibliography{schiff_moment.bib}
\end{document}